# Development of the Lymphatic System in the 4D XCAT Phantom

Roberto Fedrigo[1,2], William P. Segars[3], Patrick Martineau[4], Claire Gowdy[5],

Ingrid Bloise[1], Carlos F. Uribe[4,6], Arman Rahmim[1,2,6]*

[1] Department of Integrative Oncology, BC Cancer Research Institute, Vancouver, BC, Canada
[2] Department of Physics & Astronomy, University of British Columbia, Vancouver, BC, Canada
[3] Department of Radiology, Duke University, Durham, NC, USA
[4] Functional Imaging, BC Cancer, Vancouver, BC, Canada
[5] Department of Radiology, BC Children's Hospital, Vancouver, BC, Canada
[6] Department of Radiology, University of British Columbia, Vancouver, BC, Canada

* Corresponding Author, E-mail: arahmim@bccrc.ca

## ABSTRACT

**Purpose:** The XCAT phantom allows for highly sophisticated multimodality imaging research. It includes a complete set of organs, muscle, bone, soft tissue, while also accounting for age, sex, and body mass index (BMI), which allows phantom studies to be performed at a population scale. At the same time, the XCAT phantom does not currently include the lymphatic system, critical for evaluating bulky nodal malignancies in lymphoma. We aimed to incorporate a full lymphatic system into the XCAT phantom and to generate realistic simulated images via guidance from lymphoma patient studies.

**Methods:** A template lymphatic system was extended based on known anatomy was used to define 276 lymph nodes and corresponding vessels using non-uniform rational basis spline (NURBS) surfaces. Lymph node properties were modified using the Rhinoceros 3D viewing software. The XCAT general parameter script was used to input organ concentrations and generate binary files with uptake and attenuation information.

**Results:** Lymph nodes can be scaled, stretched, and translated within the intuitive Rhinoceros interface, to allow for realistic simulation of different lymph node pathologies. Bulky, heterogeneous PMBCL tumours were generated in the mediastinum using expanded lymph nodes. Our results suggest that optimized thresholding provides better accuracy for determining total metabolic tumour volume (TMTV) of PMBCL tumours, while the gradient method was most accurate for total lesion glycolysis (TLG).

**Conclusions:** An upgraded XCAT phantom with a fully simulated lymphatic system was created. Distributed to the research community, the XCAT phantom with the new lymphatic system has the potential of enabling studies to optimize image quality and quantitation, towards improved assessment of lymphoma including predictive modeling (e.g. improved TMTV and radiomics research).

**Keywords:** Phantoms, imaging, lymphoma

# 1. INTRODUCTION

Rapid developments in medical imaging have resulted in a wide variety of data acquisition and image generation methods available within the clinical setting[1–3]. The implementation of these methods continues to exceed the pace by which these techniques can be properly validated and optimized. Clinical trials are slow to implement and often suffer from limited sample sizes, resources and funding[4]. Delineating the ground truth for human subjects, such as the location or volume of a lesion, is fundamentally a time-consuming and challenging task to perform within research studies[4,5]. As a result, clinicians often need to choose between utilizing established but potentially less effective technology, or newer but less validated techniques, and potentially risk patient care[6]. Phantoms are routinely used in imaging research and clinical practice to assess image quality and quantitative accuracy of images[7–11]. However, few phantoms emulate the anatomical structures and heterogeneity features that are present in human subjects with a high degree of realism. Additionally, many phantoms do not provide users with the flexibility to modify anatomical structures[12–14]; thus making it difficult to represent variation between patients. This may limit the ability to account for biological variability within the patient population and reduce generalizability towards clinical applications.

Virtual Clinical Trials (VCTs) provide a time-efficient solution for investigating research questions at a population-scale[15–18]. In VCTs, the human subject is replaced with a digital phantom, imaging devices are simulated, and in certain cases, human observers are replaced with artificial observers. Like physical phantoms, digital phantoms have a known ground truth and can be used to evaluate and compare imaging devices and methods. Digital phantoms have the added opportunity to model human anatomy and physiology with an additional degree of sophistication. Voxelized phantoms[6], such as the VIP-MAN[19,20], define anatomical structures based on prior image segmentations of patients. However, this approach results in rough organ boundaries and limits the representation of patients with different anatomy or pathology. Boundary representation (BREP) phantoms[6] use advanced surface representations to model anatomical structures with higher degrees of realism. BREP uses mathematically defined surfaces to fit templates from patient segmentations. This results in realistic anatomical regions that can be easily manipulated to modify organ structure or account for temporal changes, such as patient motion due to respiration.

The 4D-extended cardiac torso (XCAT) phantom[21,22], a BREP-type phantom, allows for highly sophisticated multimodality imaging research. The XCAT phantom defines dozens of organs, accounts for tissue-type (e.g., muscle, bone, or soft-tissue), and allows users to modify demographic features such as age, sex, and body mass index (BMI), providing the opportunity for phantom studies to be performed at a population scale. Since the phantom is simulated, the attenuation and radioactivity ground-truth can be defined at a sub-voxel level, allowing for more sophisticated imaging metrics (e.g., shape, texture) to be evaluated. However, the XCAT phantom does not currently include the lymphatic system, critical for evaluating diseases such as the bulky nodal malignancies observed in lymphoma. The aim of this study is to incorporate a full lymphatic system into the XCAT phantom. We present the upgraded XCAT phantom, and, as a proof-of-concept, we generate realistic simulated PET/CT images via guidance from lymphoma patient studies. We focus on primary mediastinal B-cell lymphoma (PMBCL), though our work is also being extended to include diffuse large B-cell lymphoma (DLBCL) and easily translates to other forms of lymphoma and nodal malignancies.

PMBCL, a potentially curable form of non-Hodgkin's lymphoma, classically presents with bulky, heterogeneous tumour masses located within the mediastinum[23]. Positron Emission Tomography (PET) scans offer valuable insight into the progression of PMBCL. Baseline PMBCL tumour burden is often determined via [$^{18}$F]FDG PET/CT, while post-treatment scans are used to determine tumour response to therapy[24]. There is evidence that *quantitative* imaging metrics can enhance the prognostic value of PET and its ability to effectively guide treatment decisions. For instance, Total Metabolic Tumour Volume (TMTV) and Total Lesion Glycolysis (TLG) are well-documented predictors of therapy response and overall survival of lymphoma patients[25–28]. TMTV and TLG can be greatly impacted by the selected segmentation method, such as fixed thresholding (FT) or gradient-based segmentation. Thus, there is significant motivation to validate these segmentation methods for PMBCL such that they can be used to inform patient care.

In this work, we simulate realistic lymphoma patients using the XCAT phantom with a newly embedded lymphatic system. As example application, we model PMBCL using conglomerates of lymph nodes and vessels, though this work can be adapted to other studies and tasks involving the lymphatic system.

# 2. MATERIALS AND METHODS

## 2.A. Upgrades to XCAT Phantom

A template model for the lymphatic system was developed based on the anatomical data from the Visible Human Project of the National Library of Medicine (NLM)[29–31] as well as known anatomy. The Visible Male anatomical data from the NLM was segmented manually to define an initial model for the lymphatic vessels and nodes. The segmented vessels and nodes were fit with non-uniform rational basis spline (NURBS) surfaces using the surface lofting function within Rhinoceros (www.rhino3d.com); the software application used to create the original XCAT phantom. The initial lymphatic system was extended based on known anatomy with the guidance of an expert physician. Additional vessels and nodes were added within Rhinoceros to define a total of 276 lymph nodes and corresponding vessels (Fig. 1-a,b).

Given the template model, the multichannel large deformation diffeomorphic metric mapping (MC-LDDMM) method[32] was used to propagate the template to different XCAT anatomies. The MC-LDDMM method was used to calculate the transform from the Visible Male anatomy to the anatomy of different XCAT models. The major organs and bones, segmented from the Visible Male data, were used to create a voxelized image of its anatomy with the anatomical structures set to unique integer intensities. A similar image was created for each target XCAT anatomy, voxelizing the phantom with the same intensity values. Given the template and target images, MC-LDDMM calculates the transform between them; this transform is then applied to the model of the lymphatic system to define it in the given XCAT phantom. This method allows for the lymphatic system to be investigated on patients with different gender, weight, size, age, and other anatomical differences.

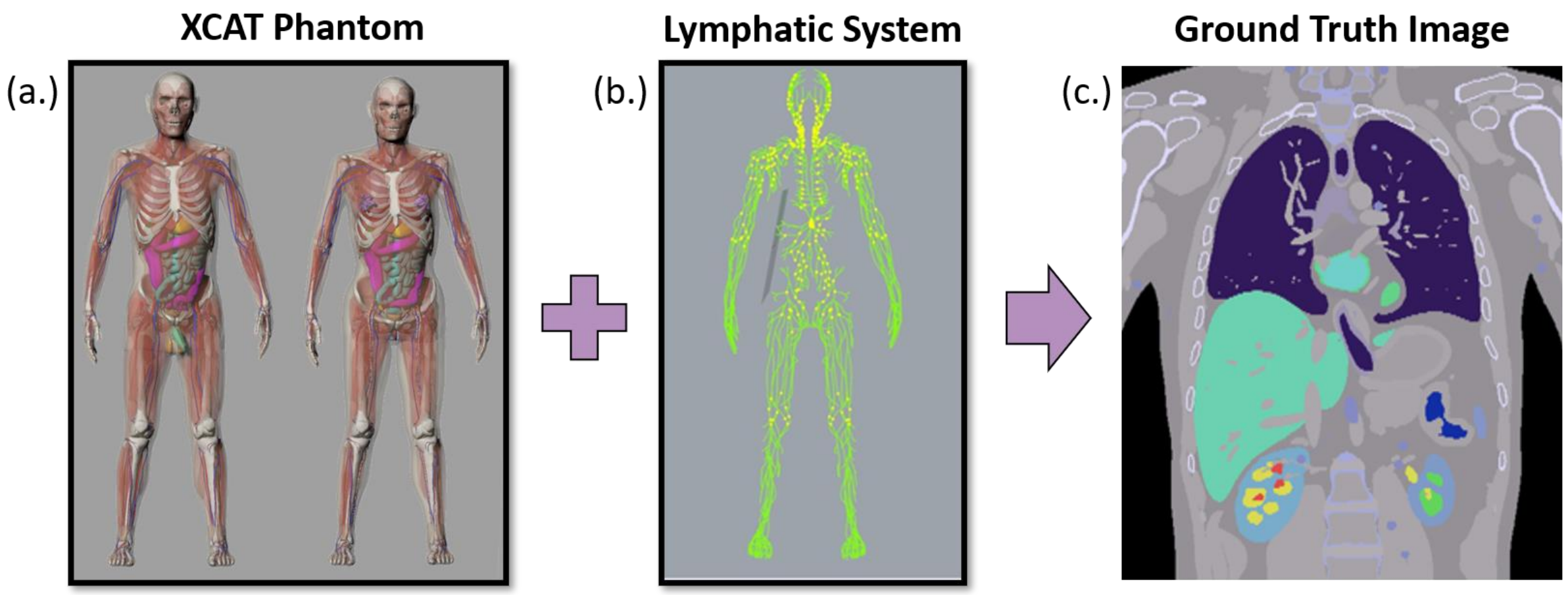

**Fig. 1:** Lymphatic system development pipeline. **(a)** Anterior views of male (left) and female (right) of 4D XCAT phantom anatomies. **(b)** Lymphatic system incorporated for XCAT phantom, which can be combined with XCAT organs. **(c)** Coronal slice of combined XCAT and lymphatic phantom showing ground truth radioactivity and attenuation distribution for a "healthy" patient.

## 2.B. Tumour Modelling and Simulation Pipeline

As an example, a pipeline was built to simulate human subjects with bulky, heterogeneous tumours located in the mediastinum. This was intended to model PET/CT images of patients with PMBCL (Fig. 2). Ten subjects were simulated using the XCAT phantom scripts[33] with our integrated lymphatic system (Fig. 1a/b): each subject had the same organ anatomy and uptake (height=176.1cm, weight=81kg), but exhibited variation in the lymph nodes. Organ activity concentrations were specified based on an analysis of [$^{18}$F]FDG PET/CT images from 5 PMBCL patients (Table 1). To create bulky lymph node conglomerates in the XCAT anatomies, lymph node properties were modified using the Rhinoceros software (Fig. 3). Lymph node morphology and function were altered: lymph nodes were expanded, stretched asymmetrically, converged within the mediastinum, and tracer concentration was increased. One lymph node conglomerate (consisting of 2-4 lymph nodes) was generated for each simulated patient, with tumour volumes ranging from 4 to 100mL. The mean radioactivity concentration (21.5 kBq/mL) corresponds to the 50$^{th}$ percentile of manually segmented lesions from a PMBCL patient analysis (22 lesions from 13 patients).

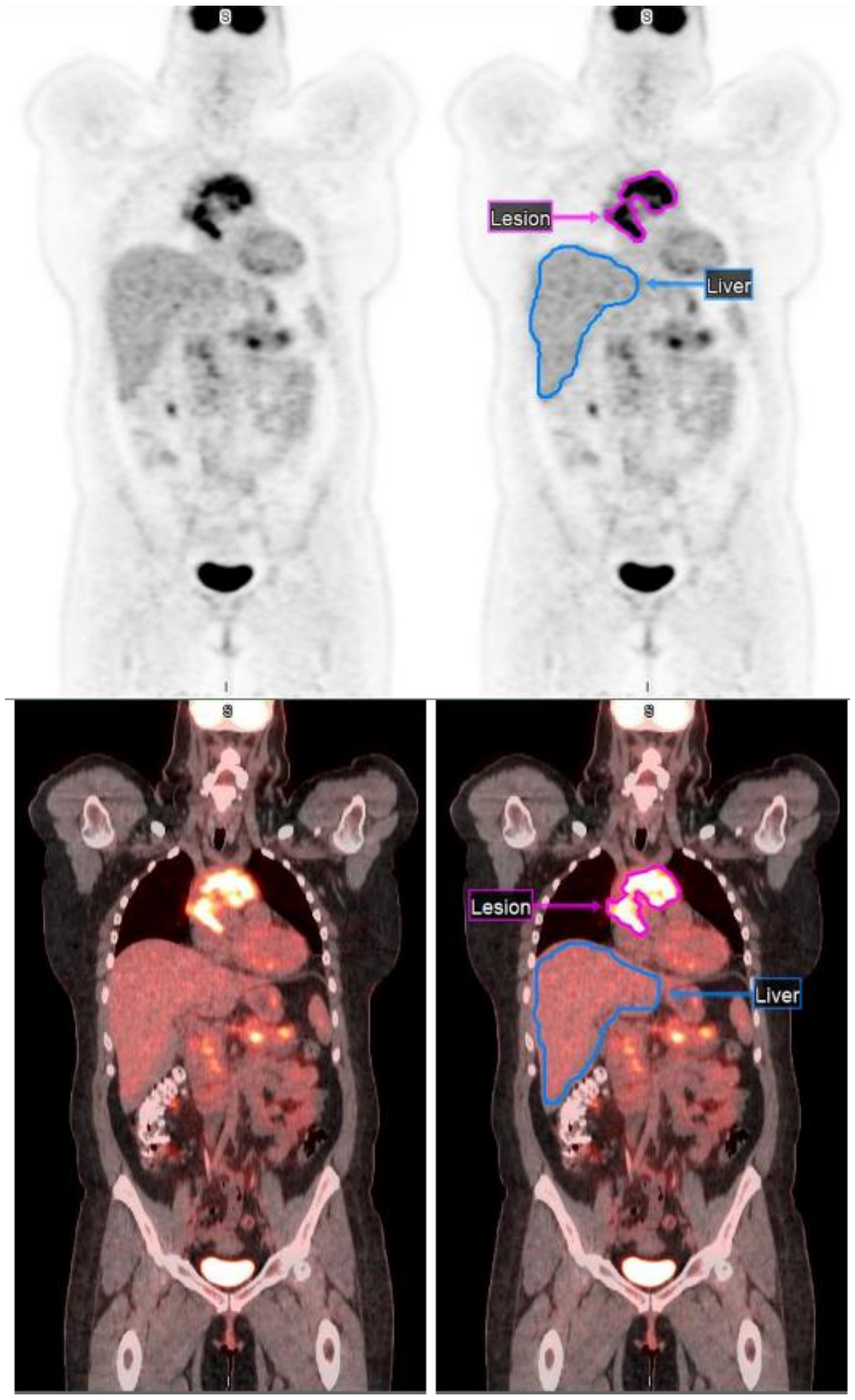

**Fig. 2:** Primary mediastinal B-cell lymphoma (PMBCL) patient with a bulky primary tumour located in the mediastinum. Lesion and liver segmentations shown in images on the right. Shown are [$^{18}$F]FDG PET (top) and [$^{18}$F]FDG PET/CT (bottom) images.

The XCAT general parameter script was used to input organ concentrations and generate binary files with ground-truth uptake and attenuation information (Fig. 1c). The phantom was used as the input to a MATLAB-based PET simulation and reconstruction tool that generates simulated PET/CT images for a GE Discovery RX scanner[34]. Ten Poisson noise-realizations were generated for each subject. Images were reconstructed with OSEM (2 iterations, 24 subsets, 256x256 matrix). Gaussian post-smoothing with different kernel sizes (2mm, 4mm, 6mm, 8mm) was applied and compared to images without post-smoothing (referred to as “Native” images). The files were converted from binary to DICOM file format and subsequently viewed using MIM (MIM Software, Inc.), a clinical radiology software program. The full image simulation pipeline and example files have been made publicly available (see data & code availability at the end of manuscript).

**Table 1:** Measured [$^{18}$F]FDG activity concentrations for PMBCL patients in different regions-of-interest.

| **Region-of-Interest** | **Concentration [kBq/mL]** |
|---|---|
| Background | 1.5 |
| Bladder | 38.5 |
| Esophagus (outer, contents) | 4.1, 3.6 |
| Heart (bloodpool, myocardium) | 11.6, 55.0 |
| Intestines | 6.7 |
| Kidney (medulla, cortex, pelvis) | 19.8, 8.4, 25.3 |
| Liver | 13.1 |
| Lung | 3.5 |
| Spleen | 7.0 |

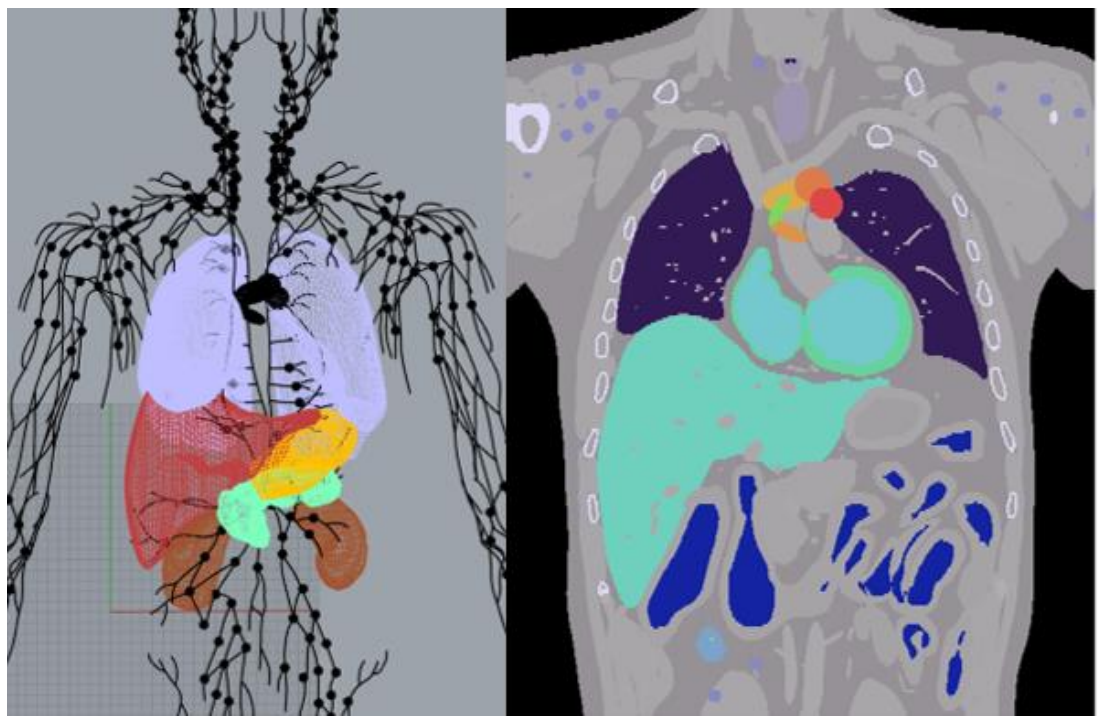

**Fig. 3:** PMBCL tumour simulation. (*Left*) Lymph nodes expanded and converged within Rhinoceros 3D viewing software, with additional organs shown. (*Right*) Coronal slice of radioactivity and attenuation file generated for patient with bulky mediastinal tumour.

## 2.C. Segmentation Analysis

Regions-of-interest (ROIs) were delineated using MIM and saved in the RTStruct file format. DICOM images and the RTStr ucts were opened in Python and the rt-utils package (https://github.com/qurit/rt-utils) was used to convert the RTStructs into binary masks. Fixed thresholding (FT) (20%, 25%, 30%, 40%, 50%) segmentations were computed using our Python code. Lesions were also segmented using MIM's gradient-based algorithm (PET Edge+). Both the FT and gradient-based tools provide time-efficient and reproducible methods for tumour segmentation. FT segmentation method sele cts voxels with concentration greater than a certain threshold (e.g. % of $SUV_{max}$). The gradient method delineates tumour edges by calculating spatial derivatives along concentration line profiles[35,36]. Total metabolic tumour volume (TMTV) and total lesion glycolysis (TLG) were calculated using each segmentation method.

To measure accuracy, percent bias (PB) for a given metric applied to a method (e.g., for TMTV or TLG applied to tumors segmented using a particular method) was defined as:

$$PB = \frac{\frac{\sum_i^n A_{measured,i}}{n} - A_{truth}}{A_{truth}} \times 100\%$$

where $A_{measured,i}$ and $A_{truth}$ are the measured and true metric values, respectively, of the $i$th noise realization. $A_{measured,i}$ is averaged over $n$ noise realizations ($n = 10$ in this study). Percent noise (coefficient of variation; COV), as a measure of precision, was in turn defined as:

$$COV = \frac{\sigma_{method}^{A}}{A_{truth}} \times 100\%$$

where $\sigma_{method}^{A}$ is the standard deviation of the $n$ measured values (e.g., TMTV or TLG) for a given subject, as obtained from multiple noise realizations of the simulated patient.

## 2.D. Patient Study

Retrospective analysis of pre-treatment PET/CT scans was performed for 14 PMBCL patients. The injected activity for each patient ranged from 281-454 MBq and PET/CT scans were performed 60min post-injection. All scans occurred between October 2005 and February 2017. Fifteen mediastinal tumours were detected by a nuclear medicine physician and delineated using 25% FT, 41% FT, gradient (PET Edge+), and manual segmentation. Percent bias and COV for TMTV and TLG was computed and compared using manual segmentation as the ground truth. Bland Altman analysis was performed, and paired t-tests were used to compare TMTV and TLG values for each segmentation method.

# 3. RESULTS

## 3.A. Simulated PET/CT Images

The lymphatic system was added to the XCAT phantom with the capability to select male or female anatomies, as well as to modify demographic features such as patient height, weight, or body mass index (BMI). To allow for realistic simulation of different lymph node pathologies, lymph nodes were scaled, asymmetrically stretched, and translated within the intuitive Rhinoceros software interface. Using this adaptable lymphatic system, lymph node conglomerates with any desired anatomical formation or tracer uptake can be simulated. As a proof-of-concept, the specific anatomy of a tumour in a PMBCL patient was replicated (Fig. 4). The generated [$^{18}$F]FDG PET images of PMBCL patients were assessed to be realistic by an experienced nuclear medicine physician. The upgraded XCAT phantom and novel application to model complex disease pathology represents a significant advancement in phantom technology.

## 3.B. Simulation Segmentations

Bulky, heterogeneous PMBCL tumours, ranging from 4mL to 100mL, were generated within the XCAT phantom mediastinum. Simulated images were visually assessed, and the 6mm post-smoothing kernel size was selected to sufficiently reduce noise in PET images. Fixed thresholding (FT) and gradient-based segmentations for the 6mm kernel size are shown in Fig. 5. The gradient-based method smoothly follows the curvature of the lesion and sharp corners are not observed. By comparison, the FT method accepts only specific values of $SUV_{max}$ based on the set percentage, and as such, jagged contours are observed due to the voxel-by-voxel nature of the segmentation method. It should be noted that certain regions within the tumour appeared to be neglected for 40-50% FT.

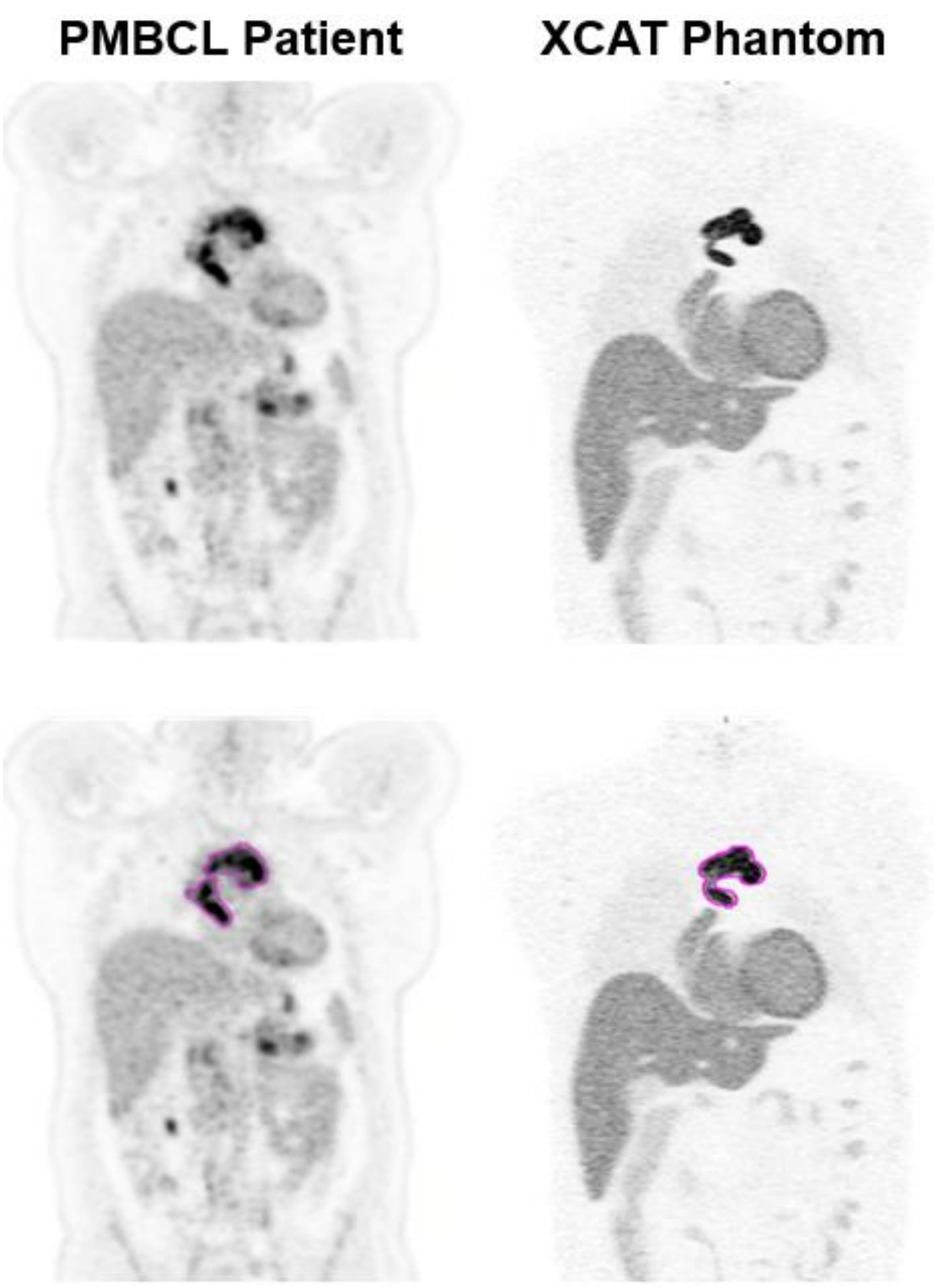

**Fig. 4:** Coronal views using MIM. Bottom images include gradient-based PET Edge+ segmentations of tumour. (*Left*) [$^{18}$F]FDG PET images of PMBCL patient with bulky mediastinal tumour. (*Right*) Simulated PET image using XCAT phantom. Tumour consists of 5 expanded lymph nodes with heterogeneous activity concentrations.

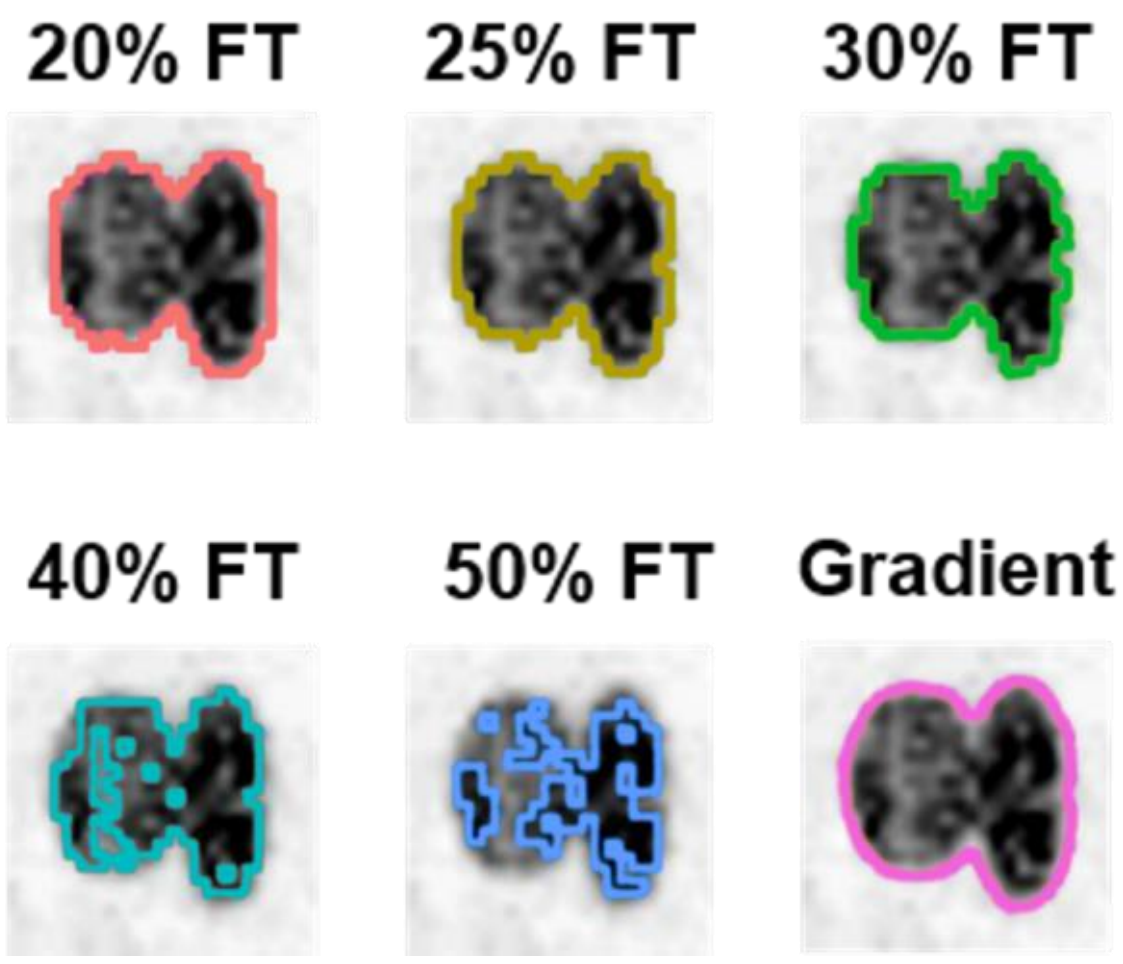

**Fig. 5:** Simulated PMBCL tumour segmented with 20%, 25%, 30%, 40%, 50% fixed threshold (FT) and gradient-based PET Edge+ method.

TMTV percent bias is plotted versus ground truth in Fig. 6. TMTV percent bias, obtained with 25% fixed thresholding, for the different tumour volu mes was 1.3% (13mL), 0.8% (39mL), -13.1% (71mL), and -11.9% (100mL). Percent bias in TMTV with PET Edge+ was 23.2% (13mL), 14.1% (39mL), 19.1% (71mL), and 22.5% (100mL). TLG percent bias versus ground truth is also plotted in Fig. 6. The 25% thresholding led to percent bias of -20.8% at 13mL and -23.6% for the 100mL lesion. TLG percent bias using gradient segmentation was -15.2% at 13mL and deviated -11.8% for the 100mL lesion. As shown in noise (COV) versus bias plots (Fig. 7), COV for TMTV was lowest for 20% FT, 25% FT, and gradient segmentation (3.60%, 4.08%, 1.35%, respectively). Similarly, COV for TLG was also minimized for 20% FT, 25% FT, and gradient (1.41%, 1.93%, 1.35%, respectively).

To investigate the effects of filtering on thresholding segmentation, images were smoothed with 2-8mm Gaussian filters, and compared to the "Native" image (without smoothing). As shown in Supplemental Fig. 1, 20-50% FT resulted in significant underestimation of TMTV in the Native, 2mm, and 4mm smoothed images. However, TMTV accuracy improved for images smoothed with 6mm or 8mm kernels, particularly if 20-30% FT are selected. TLG was consistently underestimated regardless of the selected threshold or kernel size. As shown in Supplemental Fig. 2, utilizing larger kernel sizes (6mm or 8mm) reduced percent noise for 20-30% FT.

## 3.C. Patient Study Segmentations

PMBCL tumours delineated using 25% FT, 41% FT, gradient-method, and manual segmentation are visually compared in Fig. 8. As predicted by the simulated PMBCL images, the 41% FT appears to neglect significant regions of the tumours. The 25% FT, gradient, and manual segmentation show minor deviations from each other, but appear to be quite similar overall.

Percent bias for TMTV and TLG is plotted in Fig. 9. Manual segmentation was used as the ground truth. Absolute error (MAE) for TMTV was comparable for the 25% FT and gradient segmentation (127.7±34.1% and 116.7±40.1%, respectively). MAE was largest using the 41% FT (141.6±38.9%). MAE computed for TLG was similar for the 25% FT and gradient methods (118±35.6% and 115.5±34.8%), and largest for 41% FT (130.5±40.1%).

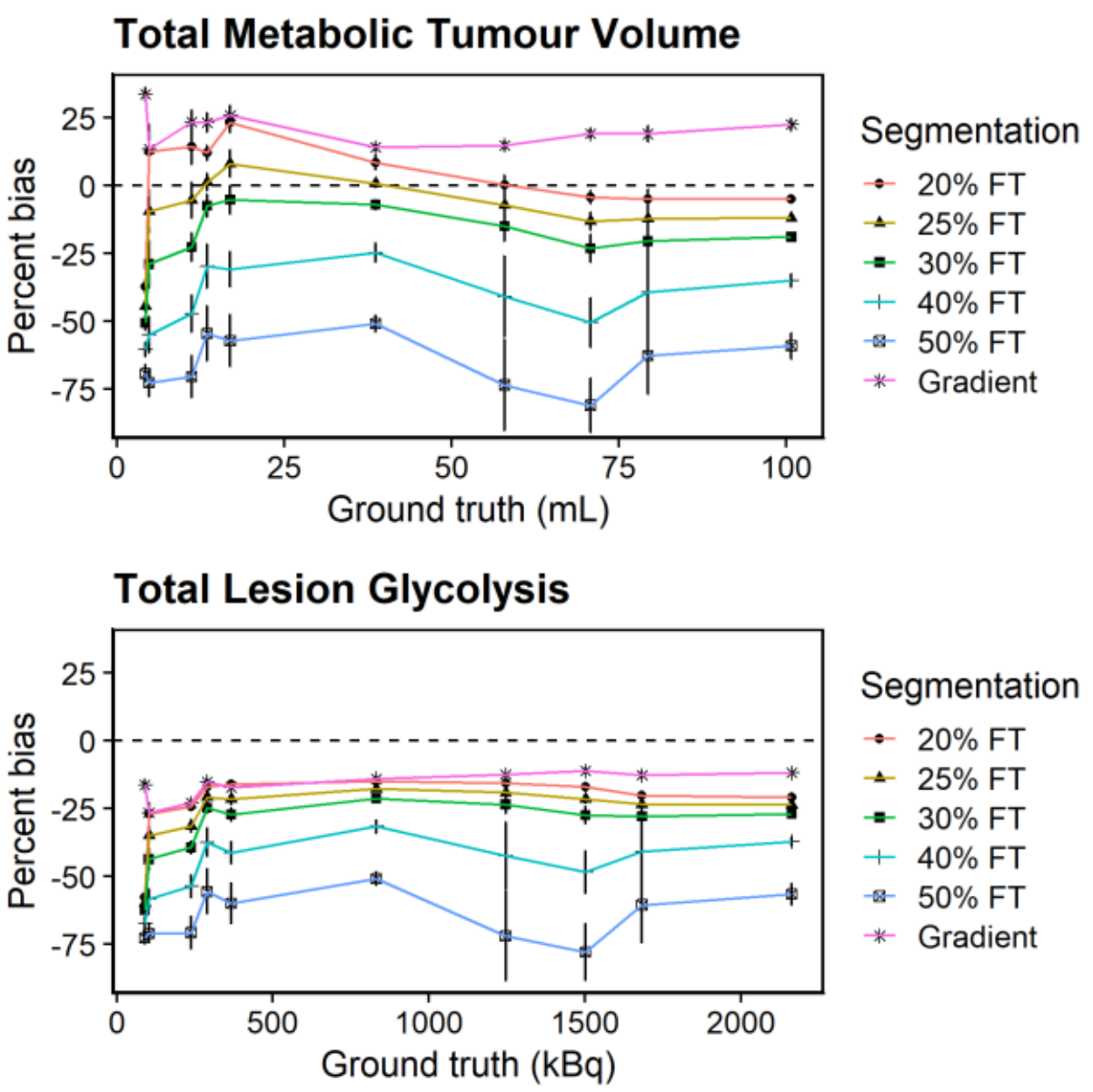

**Fig. 6:** Percent Bias+/-Percent Noise (COV) plotted versus ground truth for (*top*) TMTV and (*bottom*) TLG metrics. Each colour corresponds to either a fixed threshold (FT) or gradient-based segmentation method.

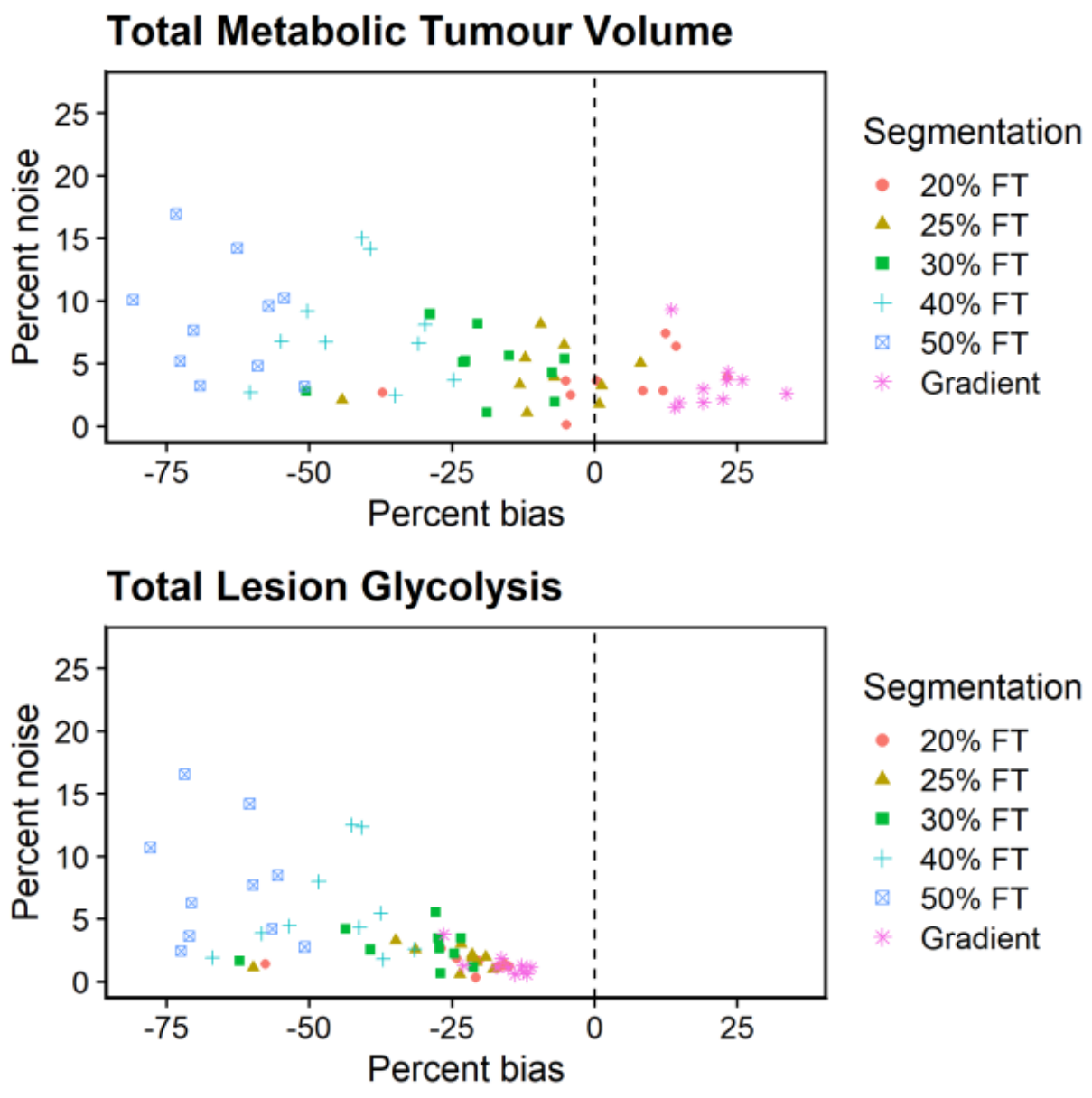

**Fig. 7:** Percent noise (COV) vs. bias plots for (*top*) TMTV and (*bottom*) TLG metrics. Each colour corresponds to either a fixed threshold (FT) or gradient-based segmentation method. Each color includes 10 points corresponding to ten different lymph node variations (ten subjects).

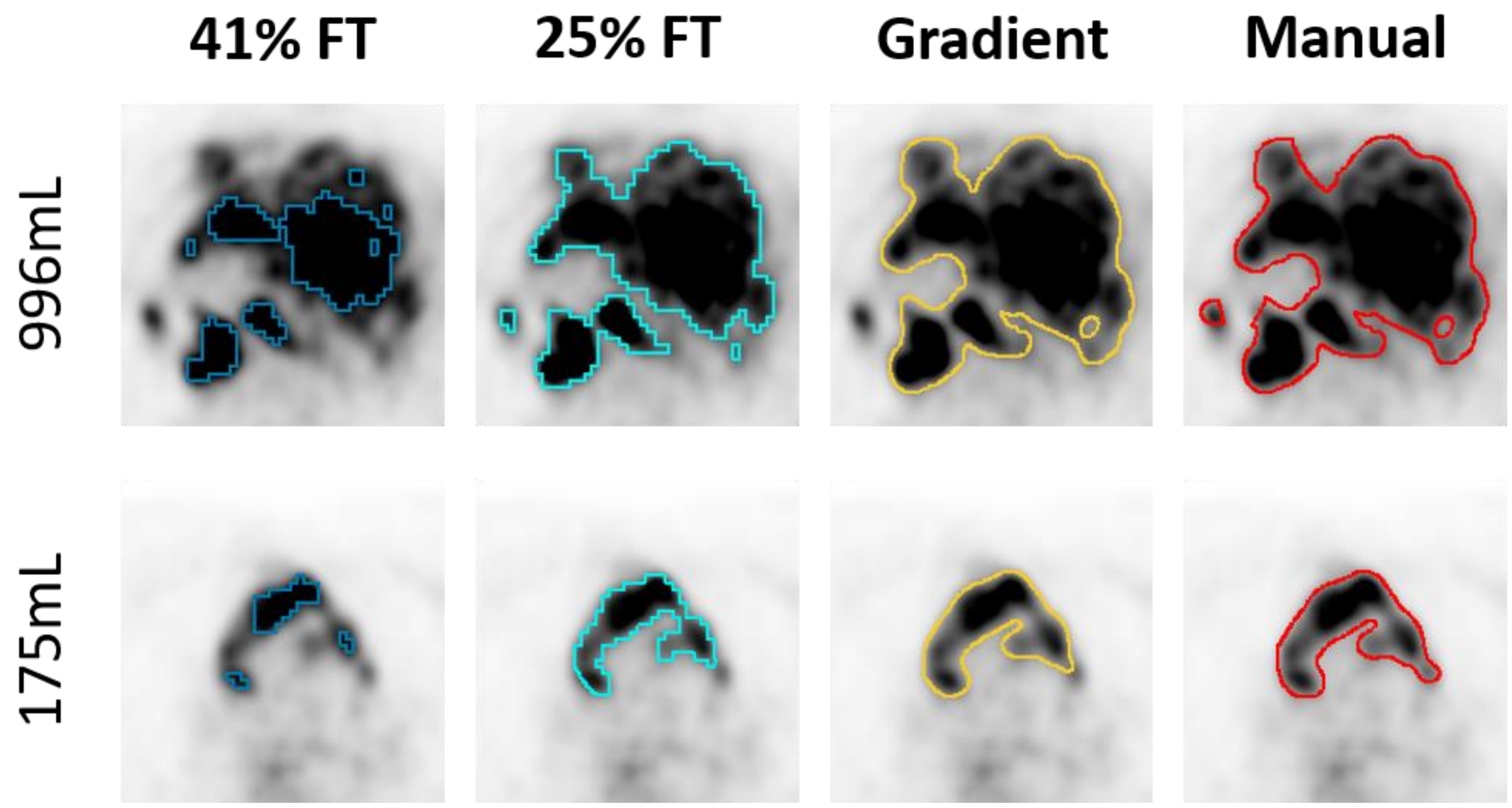

**Fig. 8:** PMBCL tumour delineated with 41% and 20% fixed threshold (FT), gradient, and manual segmentation. TMTV indicated for each tumour (determined via manual segmentation).

Percent difference between gradient and 25% FT segmentation methods is shown using Bland Altman plots (Supplemental Fig. 3). Percent Bias±Percent Noise for TMTV and TLG were 1.5±78.9% and -1.4±69.5%, with range -135.5–148.7% and -124.4–141.9%, respectively. Mean percent bias did not differ significantly from zero (P=0.944 and P=0.941). However, percent difference for larger lesions (TMTV>100mL or TLG>1000kBq) significantly differed from zero (P=3.9× $10^{-3}$ and P=0.012), and mean percent bias were 28.2±8.2% and 15.2±9.7% for TMTV and TLG, respectively.

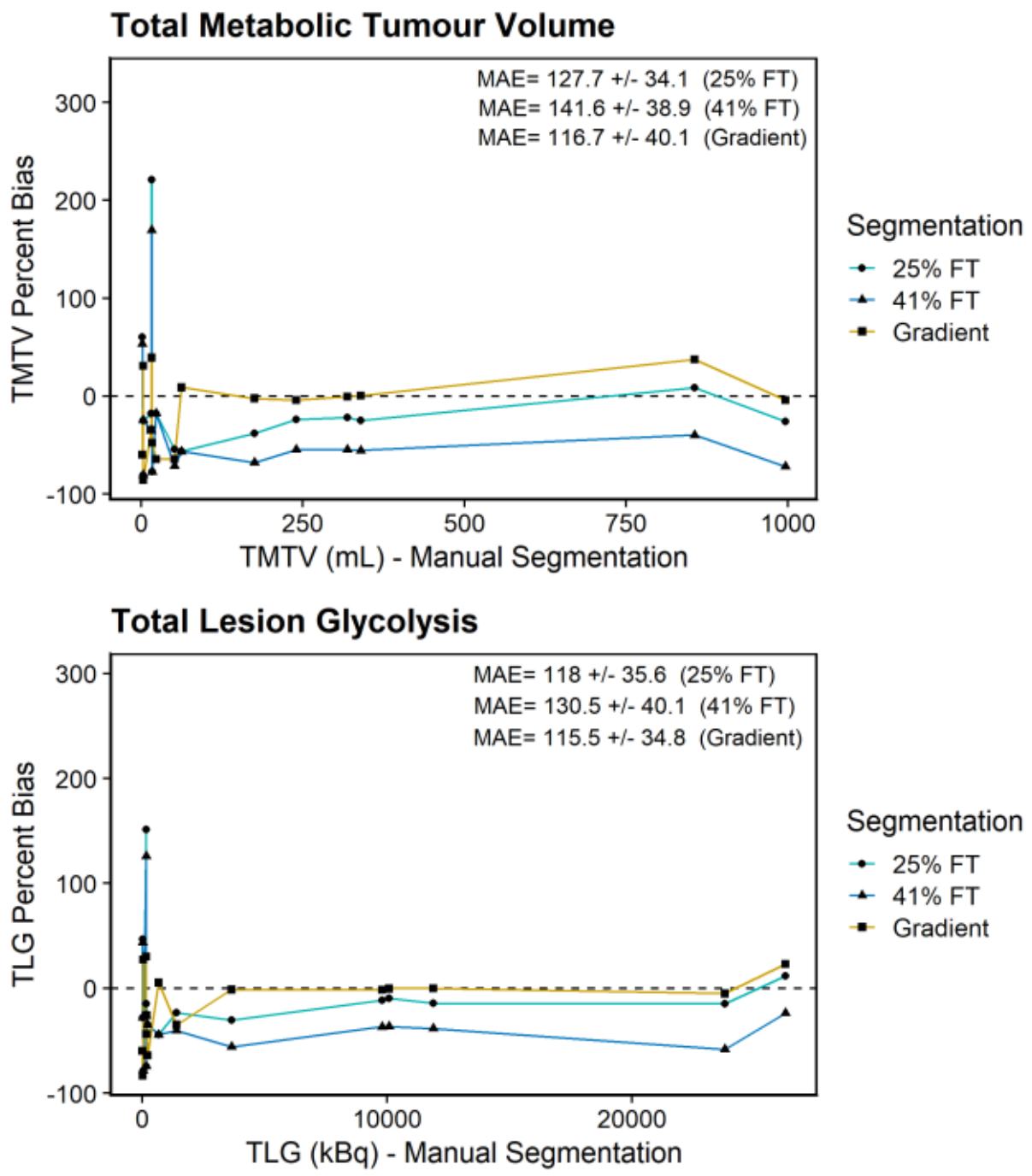

**Fig. 9:** TMTV and TLG percent bias for 25% FT, 41% FT, and gradient-based segmentation methods, using manual segmentation as the ground truth.

# 4. DISCUSSION

In this work, we upgraded the XCAT phantom to include a lymp hatic system, comprised of a network of 276 deformable lymph nodes and corresponding vessels. Through controlled manipulation of lymph node anatomy and function, lymph node pathology can be easily scaled and evaluated in digital experiments. To our knowledge, this is the most sophisticated imaging phantom to-date that models lymph nodes and vessels within a human subject. Distributed to the research community, this upgraded XCAT phantom has significant potential for improved quantitation and detectability studies in medical imaging. For instance, this provides the opportunity for VCTs in PET/CT imaging.

Conventionally, PET scanner performance is validated using the NEMA Image Quality (IQ) phantom[12,37], in which [$^{18}$F]FDG is injected into fillable spheres. The NEMA phantom, while easy to fill and reproducible, contains spheres made of plastic and the walls create "cold shells" (i.e., no radioactivity in the walls) which is not representative of real lesions. Although the NEMA phantom contains a cylindrical insert to simulate lung tissue, it neglects to model many relevant organs such as the liver, heart, or bladder (Table 2). Use of anthropomorphic phantoms, such as the Probe-IQ phantom[38–41], represents a significant framework in PET (Table 2).

**Table 2:** Comparison between standardized phantom (NEMA IQ), physical anthropomorphic (Probe-IQ), and simulated anthropomorphic (XCAT) phantom technology.

| Phantom Capabilities | NEMA IQ | Probe-IQ | XCAT (Original) | XCAT (New) |
|---|---|---|---|---|
| Lung | ✓ | ✓ | ✓ | ✓ |
| Heart | ✕ | ✓ | ✓ | ✓ |
| Skeletal Structure | ✕ | ✓ | ✓ | ✓ |
| Demographic Features | ✕ | ✕ | ✓ | ✓ |
| Lymphatic System | ✕ | ✕ | ✕ | ✓ |

Phantom inserts are used to simulate organs such as a liver, lungs, ribs, and skeletal structure (e.g., spinal cord, ribs). For example, polyurethane filter foam is used to position organs within the Probe-IQ phantom[39,42] and create small pockets of air to establish heterogeneous radioactivity concentrations, which is more representative of a real patient scan. However, physical phantoms cannot easily model complex anatomical structures or tissue types. This may reduce the generalizability of results obtained from observer studies. In general, physical phantoms are not easily modified to represent patients with anatomical defects or varying demographic features. Thus, the utility of physical phantom experiments to inform clinical practice is limited.

By defining hundreds of anatomical structures (e.g. organs, skeleton, tissue), the XCAT provides significantly improved realism compared to the Probe-IQ phantom. Since the XCAT is digitally defined, users can generate large volumes of data in a relatively short period of time. For instance, an entire patient population with lesions inserted adjacent to high-uptake organs, can be generated for tumour detection or quantitation tasks. Phantom features can also be scaled to represent varying demographic features within the patient population (Table 2). This is advantageous for VCTs that intend to evaluate diseases that have low prevalence in the patient population (e.g. PMBCL), as it is otherwise challenging to achieve statistical significance via clinical trials. To aid with this, the XCAT includes a series of models representing individuals of varying ages, heights, and weights based on patient CT data[43]. The lymphatic system can be incorporated into each of these phantoms and used as a starting point to simulate other anatomies through alteration of the XCAT anatomical scaling parameters. Another advantage of the XCAT phantom is that the ground truth is known at a sub-voxel level, allowing for more sophisticated imaging metrics (e.g., shape, texture[44]) to be evaluated. This provides the opportunity to validate more complex radiomics features within imaging, which is not currently achieved using conventional phantoms.

By developing an adaptable lymphatic system within the XCAT phantom, researchers can now try to answer questions that are more relevant to lymphoma (Table 2). The lymph nodes and vessels can be translated and scaled to represent human subjects with varying anatomy and pathology. As example application, we used the upgraded XCAT phantom to simulate patients with primary mediastinal B-cell lymphoma (PMBCL), a relatively rare form of cancer that varies significantly within the patient population. PET/CT imaging plays an integral role in diagnosing and staging PMBCL, although the accuracy of certain metrics, such as TMTV and TLG, is no t fully understood. Thus, by generating lymph node conglomerates within the mediastinum, we were able to simulate realistic PET/CT images of PMBCL patients.

Conventionally, clinical reporting of TMTV typically involves manual segmentation of lesions by a trained nuclear medicine physician (Supplemental Fig. 4). This approach is considered to be the gold-standard[26] but is time-consuming and often suffers from high intra- and inter-observer variability[45]. Alternative methods, such as no-gold-standard (NGS) methods[4,46–48], have been and are being developed for clinical cases where the ground truth is not precisely known. NGS methods rely on utilizing measures generated using multiple estimation techniques and make a number of assumptions. This is an exciting frontier and remains to be further developed and extensively studied. In any case, there is significant motivation to implement and assess semi- or fully-automated segmentation methods for use in a clinical setting such that reporting TMTV becomes a standard of care to better *tailor* therapies for each individual.

Compared with manual segmentation, fixed thresholding algorithms are efficient to implement and largely observer-independent, resulting in lower inter-observer bias[26] (Supplemental Fig. 4). Within our study, as shown by visual comparison of the simulated segmented tumours (Fig. 5), selecting a 20-30% FT appears to result in accurate delineation of the tumour. These results were validated quantitatively in Fig. 6. TMTV quantification using 25% FT was most accurate for lesions <50mL, while the 20% FT was best for lesions >50mL. All FT segmentations underestimated total lesion glycolysis. TLG error for 20-30% FTs was within 30 percent. The 40% and 50% thresholding typically had percent biases greater than 30 percent. Therefore, selecting thresholds between 20% and 30% appears to be most suitable for TMTV and TLG quantitation. Our results found that 40% thresholding was suboptimal, due to significant internal regions in the tumour not being selected in the segmentation. These findings are in agreement with those made by Sridhar et al.[36] in head and neck cancer, in which 40% and 50% FT were shown to have poor correlation with pathological tumour volume (unlike gradient based segmentation). This contrasts with the European Association of Nuclear Medicine (EANM) recommendation, which suggests 41% FT for tumour segmentation[49]. It is important to note that our results are possibly specific to the pathology of lymphoma, which characteristically exhibits heterogeneous tumour shapes and tracer uptake patterns. It should be noted that our simulation models the GE Discovery RX scanner. Care should be taken to validate these results with different image reconstruction and processing methods, and on a diverse range of scanners in the field.

Specifically for PMBCL, Meignan et al.[50] observed that 41% FT is most accurate for determining TMTV. Our contrasting results could be related to the methodology used within the phantom experiment. The phantom within our study was developed based on the pathological properties of PMBCL, which is known to exhibit tumours composed of lymph node conglomerates. Comparatively, Meignan et al.[50] modelled tumours using anatomy that has been previously observed in PET images, by injecting radioactivity into saline bags. This highlights an opportunity created by the upgraded XCAT phantom to model pathological properties of lymphoma, rather than observed anatomy.

Gradient-based segmentation methods have shown recent advancements in lesion segmentation tasks, although these methods are highly vendor-specific and need to be carefully validated prior to introduction into a clinical setting[26]. On average, the specific gradient-based segmentation tool evaluated within our study delineated larger volumes than 20% FT segmentation. In fact, the gradient-based method consistently overestimated TMTV (Fig. 6). By contrast, the gradient-based method was reasonably accurate for determining TLG. These were encouraging results, since previous studies have observed gradient-based methods to have poor performance for heterogeneous or low-uptake tumours[26], which is characteristic of PMBCL. The overestimated tumour boundaries for the gradient method partially compensated for spillover due to the partial volume effect[51]. Given the development of partial volume correction (PVC) strategies in PET[52–54], it would be interesting for future studies to evaluate the effect of PVC on TMTV and TLG quantitation.

To highlight the realism of the XCAT phantom, patient images of PMBCL patients were also evaluated for comparison purposes. As shown in Fig. 9, the 25% FT and gradient-based segmentation methods determined similar TMTV and TLG, as compared with the manual segmentation. The recommended 41% FT significantly underestimated TMTV and TLG, compared to the manual segmentation. Given these results, it appears that 41% FT may not be optimal for PMBCL segmentation tasks. Due to the high degree of variability associated with manual segmentation[26], it is important to consider that these findings may not generalize to different scanners and reconstructions. Further investigation is required to confirm whether this trend among segmentations is consistent for multiple independent observers. The relationship between 25% FT and gradient segmentation was most consistent for bulky mediastinal tumours (>100mL). The gradient method consistently estimated larger TMTV and TLG values than the 25% FT. These findings were predicted from the simulated PET/CT images. However, the percent difference for smaller tumours (<100mL) did not follow a consistent trend. The inconsistent performance for smaller tumours may be related to biological variation between patients, which is significant in PMBCL.

Variation in tumour location, shape, and tracer uptake may uniquely impact each segmentation method. For instance, cold-spots in the tumour may uniquely influence the 25% FT, which segments on a voxel-scale, while the gradient method may not be impacted as it depends on larger-scale uptake patterns within the tumour[35,36]. In agreement with the simulated experiment, gradient segmentation resulted in larger tumour volume and activity values, compared to the 25% FT. We found that total lesion glycolysis was more robust against Poisson noise than total metabolic tumour volume; however, it is also necessary to consider the biological and pathological reproducibility of each metric prior to introduction in a clinical setting. Overall, our results suggest usage of 20-25% FT for delineating tumour volume, and gradient-based segmentation for determining total lesion glycolysis. Given our paradigm, future efforts can include assessment of accuracy and reproducibility for a range of higher-order imaging (radiomics) features.

# 5. CONCLUSIONS

The digital anthropomorphic XCAT phantom was upgraded to contain a lymphatic system. The upgraded XCAT phantom has the potential to improve image quality and quantitation experiments, towards improved assessment of lymphoma such as with predictive modelling (e.g., improved radiomics research). As a proof-of-concept, PMBCL tumours were modelled, and fixed thresholding and gradient-based segmentation methods were evaluated towards optimized PET/CT quantitation of TMTV and TLG. Though the study was applied to lymphoma, this resource is also relevant to studies and virtual imaging trials that intend to evaluate the lymphatic system.

## ACKNOWLEDGMENTS

This project was in part supported by Natural Sciences and Engineering Research Council of Canada (NSERC) Discovery Grant RGPIN-2019-06467, the Canadian Institutes of Health Research (CIHR) Project Grant PJT-173231, and the National Institutes of Health (NIH) grant P41EB028744. The authors acknowledge helpful discussions with Dr. Fereshteh Yousefirizi, Dr. Ivan Klyuzhin, Dr. Don Wilson, Dr. Kerry Savage and Dr. Laurie Sehn.

## CONFLICT OF INTEREST

The authors have no relevant conflict of interest to disclose.

# DATA AND CODE AVAILABILITY

All codes (including PET simulation and reconstruction algorithms, dicom conversion scripts) and simulated datasets are shared publicly at: https://Qurit.ca/software and https://github.com/qurit/xcat-sims

The upgraded XCAT phantom is available upon request from Dr. Paul Segars: paul.segars@duke.edu and additional information can be accessed at: https://olv.duke.edu/industry-investors/available-technologies/xcat/

# Supplement

## Supplemental Figures:

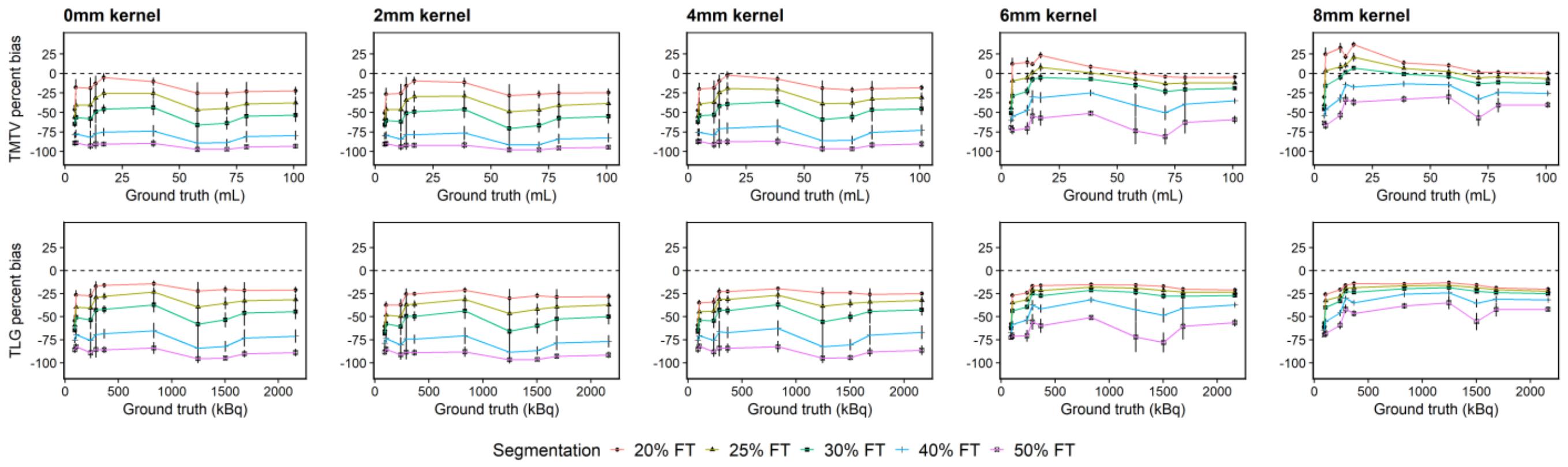

**Supplemental Figure 1:** Percent Bias±Percent Noise (COV) is plotted versus ground truth for (top) TMTV and (bottom) TLG metrics. Each column corresponds to images with different post-smoothing kernel sizes. Each colour corresponds to 20%, 25%, 30%, 40%, 50% fixed thresholding (FT) segmentation.

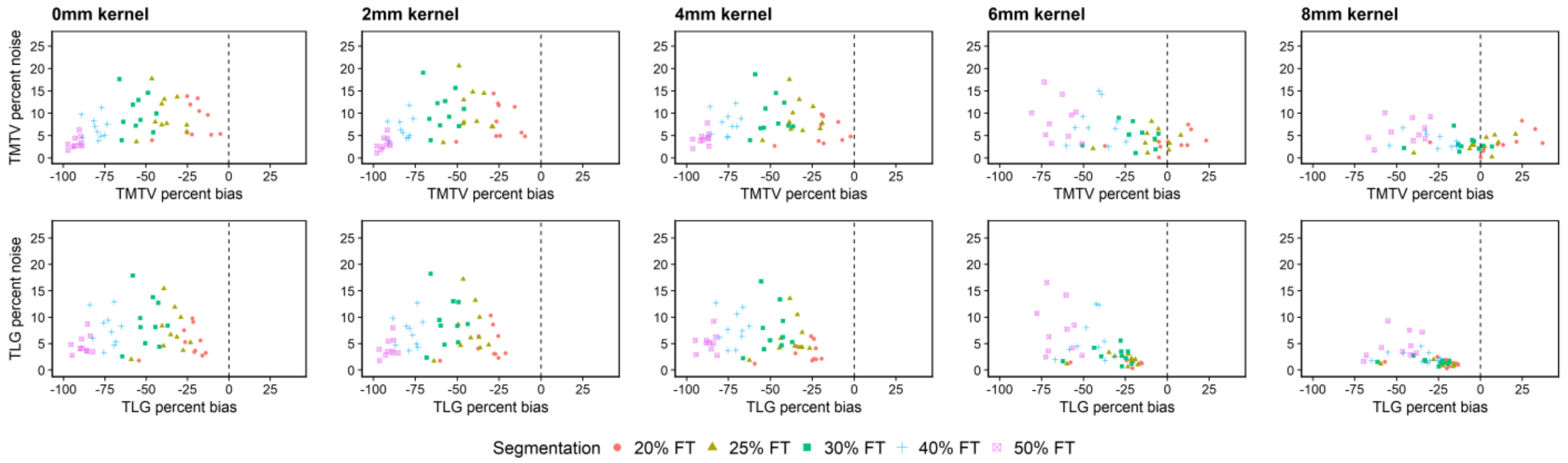

**Supplemental Figure 2:** Percent noise (COV) vs. bias plots for (top) TMTV and (bottom) TLG metrics. Each column of plots corresponds to images with different post-smoothing kernel sizes applied. Each colour corresponds to 20%, 25%, 30%, 40%, 50% fixed thresholding (FT) segmentation. Each color includes 10 points corresponding to ten different lymph node variations (ten subjects).

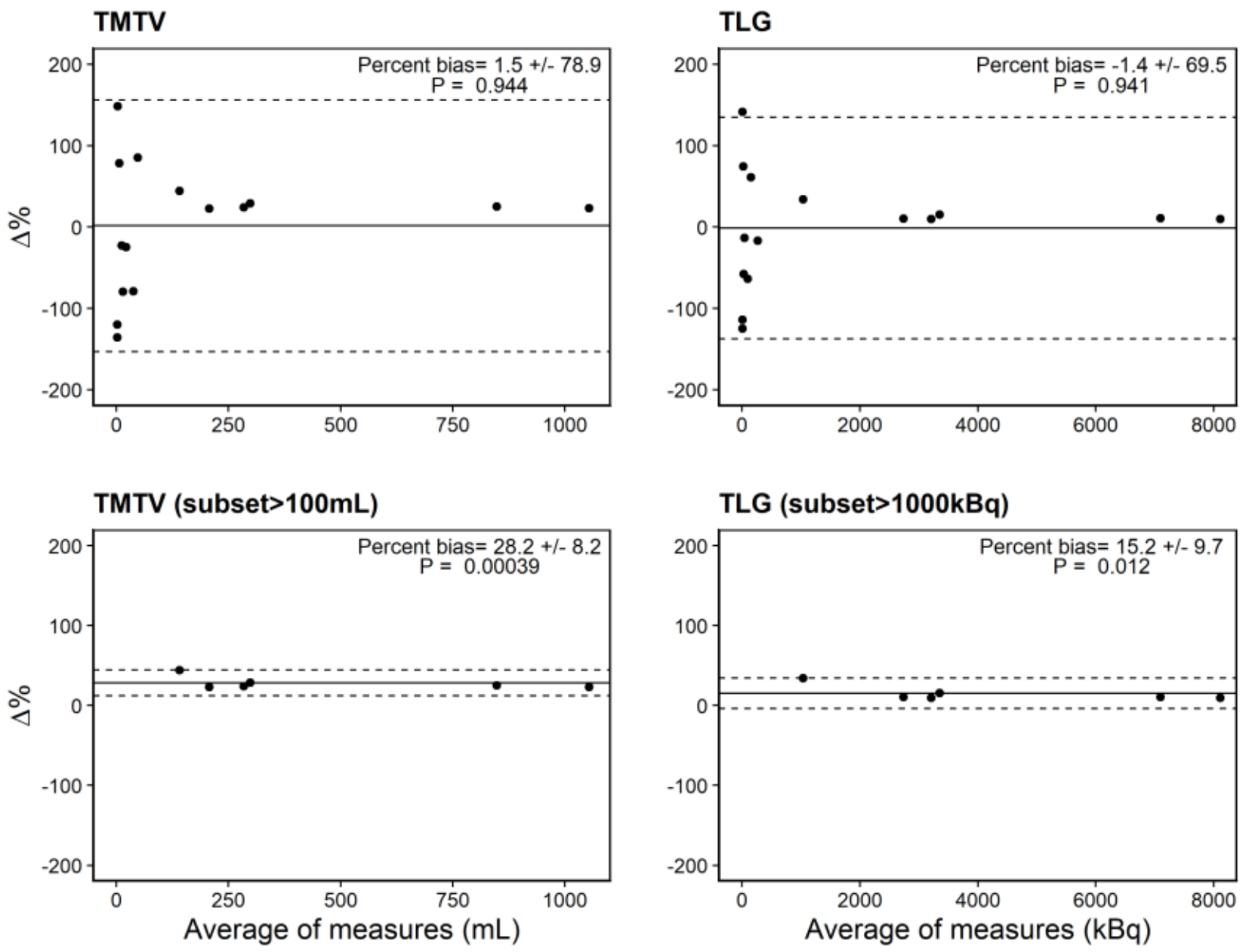

**Supplemental Figure 3:** Bland-Altman analysis plotting percent difference (Δ%) vs. average of gradient and 25% FT segmentation. TMTV and TLG are displayed for full dataset (top) and subset of data (bottom). 95% CI (dotted lines) and mean percent bias (solid lines) are shown.

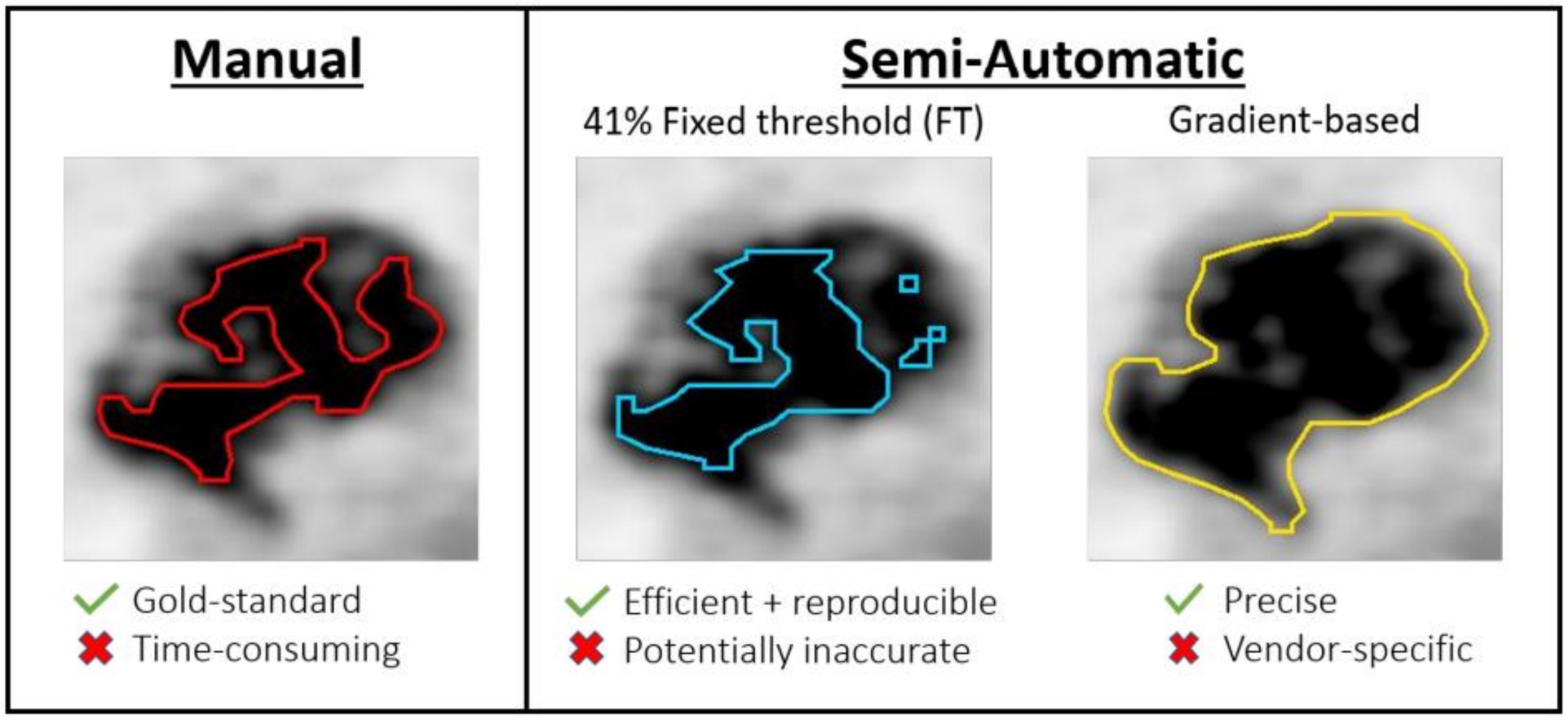

**Supplemental Figure 4:** (*Left-to-right*) Manual, fixed threshold (FT), and gradient-based segmentation methods applied to PET image of PMBCL tumour.